\documentclass{aa}

\usepackage{psfig,graphics}
\usepackage{latexsym}

\def\ros{{\sl ROSAT}}
\def\etal{{et\,al.}}

\def\degs{\ifmmode ^{\circ}\else$^{\circ}$\fi}
\def\amin{\ifmmode ^{\prime}\else$^{\prime}$\fi}
\def\asec{\ifmmode ^{\prime\prime}\else$^{\prime\prime}$\fi}

\newbox\grsign \setbox\grsign=\hbox{$>$}
\newdimen\grdimen \grdimen=\ht\grsign
\newbox\laxbox \newbox\gaxbox
\setbox\gaxbox=\hbox{\raise.5ex\hbox{$>$}\llap
     {\lower.5ex\hbox{$\sim$}}}\ht1=\grdimen\dp1=0pt
\setbox\laxbox=\hbox{\raise.5ex\hbox{$<$}\llap
     {\lower.5ex\hbox{$\sim$}}}\ht2=\grdimen\dp2=0pt
\def\gax{\mathrel{\copy\gaxbox}}

\begin{document}

\voffset=15mm

   \thesaurus{06         
              (13.07.1;  
               13.25.3)} 

   \title{Search for GRB afterglows in the ROSAT all-sky survey}

   \author{Jochen Greiner\inst{1}, Wolfgang Voges\inst{2}, 
           Thomas Boller\inst{2}, Dieter Hartmann\inst{3}}

   \offprints{J. Greiner, jgreiner@aip.de}

   \institute{Astrophysical Institute
        Potsdam, An der Sternwarte 16, 14482 Potsdam, Germany
       \and
        MPI for extraterrestrial Physics , 85740 Garching, Germany
      \and
        Clemson Univ., Dept. of Phys and Astronomy, Clemson, SC 29634-1911 USA
       }

   \date{Received 22 December 1998 / Accepted 20 May 1999}

   \authorrunning{Greiner \etal}

   \maketitle

\begin{abstract}
We report on the status of our search for X-ray afterglows of gamma-ray bursts
(GRBs) using the ROSAT all-sky survey (RASS) data. The number of potential
X-ray afterglow candidates with respect to the expected number of
beamed GRBs allows to constrain the relative beaming angles of
GRB emission and afterglow emission at about 1-5 hrs after the GRB.

\end{abstract}

\section{Introduction}

The recent discovery of long-lasting X-ray afterglow emission from (at least
some) gamma-ray burst (GRB) sources has allowed the first identification
of these enigmatic objects outside the gamma-ray range. Optical observations
of these GRB counterparts suggest a cosmological distance scale.
According to the standard scenario, the beaming angle increases with 
decreasing photon energy (Meszaros \& Rees 1997). Though it is yet unclear what
fraction of the X-ray emission is part of the GRB itself or indeed the X-ray 
afterglow produced by the deceleration of the fireball shock wave
by the interstellar medium, 
one expects that the beaming to be narrower
in the $\gamma$\-ray range, simply because $\gamma$'s come from more 
relativistic electrons and thus the 1/$\Gamma$ factor leads to more 
forward beaming.
This implies that the rate of X-ray (and other long-wavelengths) afterglows 
should be considerably higher than the GRB rate. 
A systematic search for afterglows therefore allows to constrain the beaming 
geometry of GRB emission (Rhoads 1997), in that the ratio of afterglows vs. 
GRBs determines the geometric beaming factor of the GRB emission.

\section{ROSAT all-sky survey data and expectations}

The ROSAT satellite performed the first all-sky survey in the 0.1--2.4 keV
band during 1990 August 1  -- 1991 January 25 with short additional
exposures (``repairs'') in February 16--18 and August 4--12, 1991.
During the satellites orbital period of 96 min. the telescope with a
field of view of nearly 2\degr\ diameter scans a full 360\degr\ circle
on the sky. Thus, the exposure per scan on a given sky location is
between 10--30 sec. Due to the orbital plane rotation (together with Earth's
motion) these full circles move with 1\degr/day perpendicular to the
scan direction, covering the whole sky in 6 months. Thus, a sky location 
at the ecliptic equator is covered by the telescope scans over two days,
and this coverage rises to 180 days at the ecliptic poles. Similarly,
the typical sky exposure is a function of ecliptic latitude and amounts to
200 sec at the equator and up to 40 ksec at the poles.

\begin{figure*}
    \vspace*{-0.5cm}
    \resizebox{10.5cm}{!}{\includegraphics{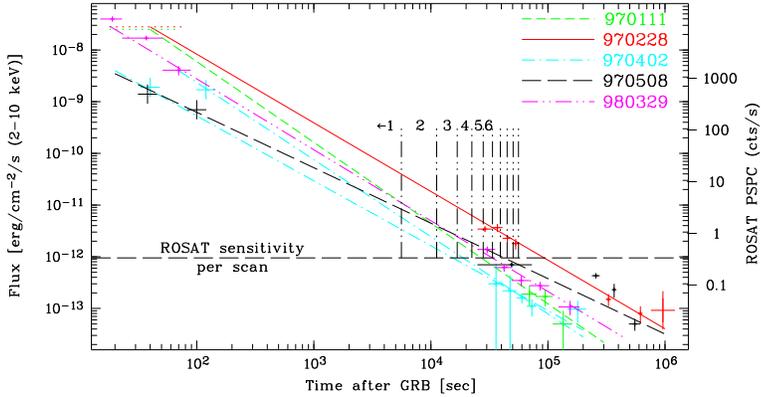}}
    \hfill
    \parbox[b]{70mm}{
    \caption[lc]{Decay light curves of some observed GRB X-ray afterglows in 
        the 2--10 keV range 
        (GRB 970111: Feroci \etal\ 1998; 
         GRB 970228: Costa \etal\ 1997;
         GRB 970402: Nicastro \etal\ 1998;
         GRB 970508: Piro \etal\ 1998;
         GRB 980329: in 't Zand \etal\ 1998)
        and their corresponding brightness extrapolated
        into the ROSAT band (scale on the right; assuming a power law with 
        photon index of --2 and
        no absorption). The horizontal line gives the sensitivity
        of the ROSAT PSPC during one scan, and the vertical lines mark
        the time windows for the possible coverage of a GRB location
        by ROSAT during its scanning mode. Thus, one may expect an afterglow
        at an intensity of up to several hundred cts/s during the first
        scan, between 0.3--8 cts/s during the second scan, 
        $<$2 cts/s during the third scan and so on.}
        }
      \label{rossen}
   \end{figure*}

ROSAT is sensitive enough to detect a
GRB X-ray afterglow for a few hours within its 10--30 sec exposure time
per sky location per scan. Fig. \ref{rossen} shows the one-scan sensitivity
of ROSAT relative to the measured X-ray afterglow decay curves.
The fraction $f$ of afterglows detectable during the RASS depends on three
critical parameters. 
First, the fraction of GRBs displaying X-ray afterglows: Previous observations
suggest this fraction to be close to one.
Second, the possible relation of X-ray flux to $\gamma$-ray peak flux or
fluence: So far, the observed X-ray afterglow fluxes at about 100 sec after
the GRB are spread within a factor of 10 only, while the GRB fluxes range
over a factor of $>$1000. 
Third, the slope of the X-ray intensity decay: Observed values range
between t$^{-1.8}$....t$^{-2.5}$. 
The effect of the combination of the latter two factors is difficult to assess
in an accurate manner given the low statistics at present, so we
base our estimate of $f$ on the observed X-ray afterglow intensities.
A comparison with the ROSAT PSPC sensitivity suggests that we would detect
practically all GRB afterglows in 3 subsequent scans, and $\sim$80\% 
in 5 scans (see Fig. \ref{rossen}). Thus, we conservatively  adopt
$f=0.8$ in the following.

The number of detectable X-ray afterglows from a GRB beamed towards us 
(based on the BATSE detection rate)  during the RASS is
$$ f \times S_R^{agl} \times R_{GRB} $$
where $R_{GRB}$ is the rate of beamed GRBs per time and area on the sky
and $S_R^{agl}$ is the effectiveness of ROSAT for afterglows in units of
time$\times$area. We adopt 
$R_{GRB}$ = 900 GRBs/sky/yr = 1 GRB/(16628$\Box \degs \times$ days).
$S_R^{agl}$ would be 122296.5 $\Box \degs \times$ days for a 100\% perfect
survey. With a temporal completeness of the RASS of
62.5\% we use $S_R^{agl}$ = 76435  $\Box \degs \times$ days in the following.
Thus, we would expect 4.6$\times$$f$ afterglows of beamed GRBs to be detected
during the RASS.

\section{Data selection and Results}

We have first produced scan-to-scan light curves for all RASS sources
with either a count rate larger than 0.05 cts/s or a detection likelihood
larger than 10, resulting in a total of 25176 light curves. 
(Note that these criteria correspond to a lower sensitivity
threshold as compared to the RASS Bright Source Catalog which invoked
at least 15 counts and a detection likelihood $\gax$ 15).
Each of these light curves consists of about 20--450 bins spaced at 96 min.,
with each bin corresponding to 10--30 seconds exposure time.

After ignoring 363 light curves with negative mean count rates (caused
by incorrect background-subtraction in our automatic procedure) 
we apply two conditions to these light curves: 
(1) the maximum bin should have a signal-to-noise ratio S/N$>$3 above the 
   mean count rate around the maximum (S/N is defined as difference between the
   maximum and mean count rate divided by the square root of the quadratic
   sum of the error of the maximum and mean count rates, respectively), and
(2) the mean count rate until one bin before the maximum as well as the
   mean count rate for times later than 5 bins after the maximum
  should be consistent with zero.
This yields 32 candidate sources.
We then excluded visually
(i) sources with double and multipeak structures (4), and
(ii) sources with a rise over 2--3 bins and zero flux immediately after the 
    maximum, e.g. like an inverted decay (3).
In addition, we investigated the available pointed observations for
3 of these candidates, and excluded all 3 sources because we find persistent 
X-ray emission at a level below the RASS threshold.
Finally, we made a correlation with various optical, infrared and radio
catalogs and excluded sources with known counterparts as well as sources 
with likely counterparts (3).
After this procedure we are left with 19 GRB afterglow candidates.

\section{Discussion}

Half of these remaining light curves are single peak events,
i.e. have just one bin with S/N$>$3 and otherwise zero count rate.
These events could be either statistical fluctuations, flare stars, or
GRB afterglows with a very fast decay slope. 
We note that these single peak events cannot be distinguished from late-type
flare stars other than by optical follow-up investigations, since typical time
scales of flares range between 10--60 min., just as we observe.
We therefore divide our conclusion into two cases:
\begin{enumerate}
\vspace{-0.2cm}\item
If the single peak events turn out to be due predominantly to flare stars, 
then the remaining 9 multipeak events (and light curves consistent with a
power-law decay) found versus 4 events expected is at first order consistent 
with the BeppoSAX picture and implies that beaming may not be required.
\item If the single peak events are GRB afterglows then geometric beaming of
the GRB emission is possible, but at a moderate value.
\vspace{-0.2cm}
\end{enumerate}

In order to more accurately determine the fraction $f$ and thus the expected
number of events, a distribution function of power-law indices of GRB 
X-ray afterglow decays as measured by BeppoSAX is highly needed.

We finally note that GRBs from directions within a few degrees of the galactic 
plane will suffer absorption, thus compromising the sensitivity of the
ROSAT detection. However, this effect concerns only a few percent of the sky,
and mainly applies to single peak events, because any
multi-peak event should have a first peak count rate high enough to 
still be detectable.

\begin{acknowledgements}
We are indebted to E. Costa and J. in 't Zand for providing X-ray afterglow 
light curves in digital form.
JG is supported by the German Bun\-des\-mi\-ni\-sterium f\"ur Bildung,
Wissenschaft, Forschung und Technologie
(BMBF/DLR) under contract No. 50\,OR\,96\,09\,8.
The \ros\, project is supported by BMBF/DLR and the Max-Planck-Society.
This research has made use of the Simbad database, operated at CDS, 
Strasbourg, France and the Digitized Sky Survey (DSS) produced at 
the Space Telescope Science Institute under US Government grant NAG W-2166.
\end{acknowledgements}


\begin{thebibliography}{}

\bibitem[]{cfh97} Costa E., Frontera F., Heise J., \etal\
  1997, Nat. 387, 783

\bibitem[]{fag98} Feroci M., Antonelli L.A., Guainazzi M., \etal\ 1998, 
   A\&A 332, L29

\bibitem[]{zaa98} in 't Zand J.J.M., Amati L., Antonelli L.A., \etal\
  1998, ApJ 505, L119

\bibitem[]{mr97} Meszaros P., Rees M.J., 1997, ApJ 476, 232

\bibitem[]{naab98} Nicastro L., Amati L., Antonelli L.A., \etal\
  1998, A\&A 338, L17

\bibitem[]{paa98} Piro L., Amati L., Antonelli L.A., \etal\ 1998, A\&A 331, L41

\bibitem[]{rhoad97} Rhoads J.E., 1997, ApJ 487, L1

\end{thebibliography}
\end{document}